\RequirePackage{fix-cm}
\documentclass[smallextended]{svjour3}       
\smartqed  
\usepackage{graphicx, cite}
\begin{document}

\title{Photon modulated coherent states of a generalized isotonic oscillator by Weyl ordering and their non-classical properties }


\author{V. Chithiika Ruby        \and
        M. Senthilvelan 
}


\institute{ Centre for Nonlinear Dynamics, Bharathidasan University, Tiruchirappalli -- 620024, Tamilnadu, India\\
              \email{velan@cnld.bdu.ac.in} }

\maketitle

\begin{abstract}
We construct photon modulated coherent states of a generalized isotonic oscillator by 
expanding the newly introduced superposed operator through Weyl ordering method. 
We evaluate the parameter $A_3$ and the $s$-parameterized
quasi probability distribution function to confirm the non - classical nature of the 
states. We also calculate the identities related with the quadrature squeezing to 
explore the squeezing property of the states. Finally, we investigate the fidelity between 
the photon modulated coherent states and coherent states to quantify 
the non-Gaussianity of the states. The constructed states and their associated non - classical 
properties will add further knowledge on the potential. 
\end{abstract}

\section{Introduction}
In quantum optics, the ordering of non-commuting operators,
say annihilation ($\hat{a}$) and creation $(\hat{a}^{\dagger})$ operators,
with $[\hat{a}, \hat{a}^{\dagger}] = 1$,
play a central role since most of the physical quantities are calculated through
the expectation values of various  operator valued functions of these two non-commuting operators.
The rules which are being widely used to express these two
operators in ordered form are (i) normal, (ii) antinormal and (iii) Weyl ordering \cite{agar}. 
Unlike the normal/antinormal ordering in  the Weyl ordering one enlists all
possible combinations of these two operators \cite{agar}.
These operators perform the action of subtraction/addition  of a photon from/to
a field and are represented  by the equations $\hat{a}|n\rangle = \sqrt{n}|n-1\rangle$
and $\hat{a}^{\dagger} |n\rangle = \sqrt{n+1}|n+1\rangle$ respectively.
To explore the annihilation and creation operators, which obey 
the relation $[\hat{a}, \hat{a}^{\dagger}] = 1$, of
 some exactly solvable potentials  that
possess linear energy spectrum (besides the harmonic oscillator), one may 
utilize the shape invariance property \cite{bal, chi3, chi4, chi5}.

The photon added coherent state was first introduced as an intermediate state
by Agarwal and Tara \cite{agarwal1}. This coherent state can be obtained  by repeatedly operating  
the  photon creation operator on a coherent state. The resultant state is shown to
exhibit non-zero amplitude and admits certain non-classical properties such as phase
squeezing and sub-Poissonian statistics. Such an intermediate state can be experimentally realized
by the passage of two-level atoms, which are kept in the excited state, through a  cavity that
encloses an $m$-photon medium \cite{agarwal1}.
The photon added coherent state can also be generated through the parametric down conversion
process \cite{agarwal1}. 
The photon subtraction from a field has been demonstrated experimentally by Wenger
and his coworkers by observing  some photons which split from the initial field when 
it passes through a beam splitter of high transmittivity \cite{grain}.
The single-photon addition to the coherent field using a `non-degenerate parametric down-converter' has
also been demonstrated  recently \cite{exp_pm}. It is clear from the experimental evidences as
well as theoretical arguments that the photon addition and subtraction
can be used as probes to generate various non-classical
states.  The photon addition and subtraction are  important tools to test certain 
fundamental quantum phenomena. The experimental generation of the photon added and subtracted states helps us
to verify quantitatively the bosonic commutation relation between the 
annihilation and creation operators, that is $[\hat{a}, \hat{a}^{\dagger}]  =1$ \cite{kim1}. 
Besides this, the photon addition and subtraction can
be applied to improve entanglement between Gaussian states \cite{gaussian_state}. 
They also serve  as a promising tool for quantum-state  
engineering \cite{qeng}. 

Inspired by the wide number of applications, in recent years, attempts have also 
been made to construct new quantum states through the elementary action of 
coherent superposition of photon addition and subtraction operators, that is
$\mu \hat{a}+\nu \hat{a}^{\dagger}$, to act upon a coherent state \cite{leen, chatr}.  
In this direction, very recently, Xu et al have introduced a new generalized photon modulated thermal state 
by considering the generalized superposed operator $(\mu \hat{a}+\nu \hat{a}^{\dagger})^N$  \cite{xu}.
The authors have derived normal ordering  form of $(\mu \hat{a}+\nu \hat{a}^{\dagger})^N$ using the
generating function of a Hermite polynomial. In this paper, we consider this generalized 
superposed operator and expand it through Weyl ordering. We then obtain an explicit
expression of this superposed operator in terms of $:\hat{a}^{\dagger^m}\hat{a}^n:$, 
where the symbol $: :$ denotes normal ordered product form of bosonic operators and 
$m$ and $n$ are arbitrary integers. We use this expanded form of the generalized operator to
construct the photon modulated coherent states of the generalized isotonic oscillator potential (\ref{pot}). 
We investigate the non-classical properties of the photon
modulated states by evaluating  the parameter $A_3$ \cite{agar1},
quadrature squeezing \cite{walls} and $s$-parameterized function \cite{cah}. Our
results confirm the non-classical nature and squeezing property of the photon modulated
coherent states. The photon modulated coherent states constructed in this paper to the potential (\ref{pot}) is 
new and add additional results to this system. 

The photon modulated coherent states can be generated experimentally through the interference 
between the coherent and Fock states using  ``quantum catalysis" \cite{qcat}. In a recently proposed 
 quantum catalysis technique  \cite{tim}, the interference is acheived through the variable beam splitter 
that comprises a polarizing beam splitter, a half-wave plate and interference beam splitter. 
Though this technique,  the probability amplitude of a coherent state is 
modulated by means of photons (Fock states). Hence, the conditional preparation of this kind of multi-photon states 
shows the tunability  into the  non-classical regime which finds applications in quantum metrology,  computation and communication \cite{tim}. Motivated by this experimental progress, we intend to construct photon modulated coherent states for the generalized isotonic 
oscillator. 

In the following section, we recall the construction of coherent states for the 
generalized isotonic oscillator. In Sec. 3, we express the superposed operator $(\mu \hat{a} + \nu \hat{a}^{\dagger})^N$ through
Weyl ordering. We construct the photon modulated coherent states for the
generalized isotonic oscillator in Sec. 4. In Sec. 5, we evaluate the parameter $A_3$,
quadrature squeezing identities, $s$-parameterized quasi-probability distribution function
and the fidelity. Finally, we summarize our results in Sec. 6.

\section{Generalized isotonic oscillator}
Very recently we have constructed the annihilation and creation operators satisfying the Heisenberg-Weyl
algebra for the following generalized isotonic oscillator potential
\cite{chi3, chi4} 
\begin{equation}
V(x) =  x^2 + 8 \frac{(2 x^2 - 1)}{(2 x^2 + 1)^2}.
\label{pot}
\end{equation}
The Schr\"{o}dinger equation associated with this potential admits the following eigenfunctions and
eigenvalues \cite{car},
\begin{eqnarray}
\psi_n(x) &=& N_n \displaystyle{\frac{{\cal P}_n(x)}{(1+2x^2)}} e^{-x^2/2}, \label{eig}\\
E_n &=& -\frac{3}{2} + n, \quad n = 0,3,4,...\label{gk3}
\end{eqnarray}
respectively, where the normalization constant is given by
\begin{eqnarray}
N_n =\left[\displaystyle{\frac{(n-1)(n-2)}{2^n n! \sqrt{\pi}}}\right]^{1/2}, \quad n = 0,3,4,... .
\label{gk4}
\end{eqnarray}
The newly defined ${\cal P}$-Hermite polynomials  are given by
\begin{eqnarray}
\hspace{-1cm} \qquad \quad {\cal P}_n(x) = \left\{\begin{array}{c}
               1, \hspace{5.5cm} \;\mbox{if}\;\;n = 0,\\
                H_n(x) + 4 n H_{n-2}(x) + 4 n (n-3) H_{n-4}(x), \;\mbox{if}\;\;n = 3, 4, 5,...\;.\\
                       \end{array}\right.
\end{eqnarray}
While deriving the expressions (\ref{eig}) and (\ref{gk3}),
Planck's constant ($\hbar$) and the mass ($m_0$) are
absorbed suitably.

In our earlier studies, besides the number states (\ref{eig}), we have
shown that the potential (\ref{pot}) admits coherent states and various
non-classical states including nonlinear coherent states and nonlinear squeezed states
\cite{chi3, chi4, chi2}. While exploring the ladder operators from the recurrence relations, involving the 
eigenfunctions (\ref{eig}), we obtained them only as $f$-deformed ones \cite{chi3}. In other words,
these deformed operators act on the number states to yield
\begin{eqnarray}
\hat{N}_{-}|n\rangle &=& \sqrt{n}\;f(n)\;|n-1\rangle, \label{lada} \\
\hat{N}_{+}|n\rangle &=& \sqrt{n+1}\;f(n+1)\; |n+1\rangle,
\label{lad10}
\end{eqnarray}
with $f(n) = \sqrt{(n-1)(n-3)}$. We considered these operators as $f$-deformed partner
of the generalized isotonic oscillator (\ref{pot}). By transforming them into new ones, namely
\begin{eqnarray}
\hspace{1cm}\hat{a} &=& \sqrt{\frac{\hat{n} - 2}{\hat{N}_{-}\hat{N}_{+}}}\hat{N}_{-}, \;\; \qquad \hat{a}^{\dagger} = \hat{N}_{+}
\sqrt{\frac{\hat{n} - 2}{\hat{N_{-}}\hat{N}_{+}}},
\label{newop}
\end{eqnarray}
so that the new ones are self-adjoint to each other and their combination satisfy the Heisenberg-Weyl algebra 
(for more details one may refer Refs. \cite{chi3, chi4}), that is
\begin{eqnarray}
\hspace{-1cm} \qquad [\hat{a}, \hat{a}^{\dagger}]|n\rangle &=& |n\rangle, \;\; [\hat{a}^{\dagger}\hat{a}, \hat{a}]|n\rangle = -\hat{a}|n\rangle, \;\; \;[\hat{a}^{\dagger}\hat{a}, \hat{a}^{\dagger}]|n\rangle = \hat{a}^{\dagger}|n\rangle.  \label{he3}
\end{eqnarray}
We then proved that  the new operators factorize the Hamiltonian $\hat{H'} = \hat{H} + E_0$ as
$\hat{H}  =  \hat{a}^{\dagger}\hat{a}$,
where $\hat{H}$ is the Hamiltonian associated with the potential (\ref{pot}).

In this way we have obtained  the annihilation and creation operators for the
generalized isotonic oscillator potential (\ref{pot}) that possesses linear energy spectrum
as in the case of harmonic oscillator. We have also
obtained the coherent states,
\begin{equation}
|\zeta \rangle = e^{-|\zeta|^2/2} \sum^{\infty}_{n = 0} \frac{\zeta^n}{\sqrt{n!}} |n+3\rangle.
\label{ccs}
\end{equation}
In Sec. 4, we consider this coherent state, $|\zeta\rangle$, as input state to 
construct photon modulated coherent state. 
 
\section{  Weyl-ordered form of $(\mu \hat{a} + \nu \hat{a}^{\dagger})^N$}
\label{secs2}
We construct photon modulated coherent states of the generalized isotonic oscillator potential (\ref{pot})
by expanding the superposed operator $(\mu \hat{a} + \nu \hat{a}^{\dagger})^N$ through Weyl ordering method. 
Here $\mu$ and $\nu$ are complex parameters and $N$ is a real integer.  We then 
obtain an explicit expression of this superposed operator in
terms of $:\hat{a}^{\dagger^m} \hat{a}^n:$, where the symbol
$::$ denotes normal ordered product of bosonic operators and $m$ and $n$ are real integers.  
We mention here that this generalized superposed operator 
$(\mu \hat{a} + \nu \hat{a}^{\dagger})^N$  is not a Hermitian. 

To begin with, we expand this function in terms of power series \cite{book}, that is 
\begin{equation}
\left[(\mu\hat{a} + \nu \hat{a}^{\dagger})^N\right]_{W} = \sum^{N}_{k = 0}
                                                          \left(\begin{array}{c}
                                                         N\\
                                                         k
                                                   \end{array}\right)
\left[\left(\mu \hat{a}\right)^{k} \left(\nu \hat{a}^{\dagger}\right)^{N-k}\right]_{W},
\label{op1}
\end{equation}
where $W$ denotes Weyl ordering and $N$ is an integer. Recalling the 
Weyl ordering method for the operator $\left(\hat{a}^{n} \hat{a}^{{\dagger}^m} \right)_W$  \cite{agar1} 
\begin{equation}
\left(\hat{a}^{n} \hat{a}^{{\dagger}^m} \right)_W  =   (-1)^{m} :{\cal H}_{m,n}(z, -z^*):
\end{equation}
where
\begin{equation}
{\cal H}_{m,n}(z, -z^*) = m!\;n! \sum^{min(m, n)}_{k = 0} \frac{\left(\frac{-1}{2}\right)^k (-z^*)^{m-k} z^{n-k}}{k! (m - k)! (n - k)!},
\label{op2}
\end{equation}
is the two parameter Hermite polynomial and $z$ is the coherent state eigenvalue.  
Equation (\ref{op1}) can now be expanded to yield 
\begin{equation}
\hspace{-1cm} \qquad \qquad \left[(\mu\hat{a} + \nu \hat{a}^{\dagger})^N\right]_{W} = \nu^N N! \sum^{N}_{k = 0} \left(\frac{\mu}{\nu}\right)^k
\sum^{min(N-k, k)}_{l = 0} \frac{\left(\frac{1}{2}\right)^l:\hat{a}^{\dagger^{N-k-l}}\hat{a}^{k-l}:}{l!\; (k - l)!\;(N-k-l)!}.
\label{op3}
\end{equation}
We have derived an expression which is stated in terms of normal ordering
$:\hat{a}^{\dagger}\hat{a}:$. We consider (\ref{op3}) as the desired expression because the physical quantities 
can now be expressed in terms of the coherent state eigenvalues. Using this 
expression, we  construct the photon modulated coherent states  
of (\ref{pot}) for the coherent state $|\zeta\rangle$ given in  (\ref{ccs}).

\section{Photon modulated coherent states}
 Photon modulated coherent states of the generalized isotonic oscillator can be obtained
by letting the operator $\left(\mu \hat{a}  + \nu \hat{a}^{\dagger} \right)^{N}$,  given in (\ref{op3}), 
to act on its coherent states, that is
\begin{eqnarray}
\hspace{-1.1cm} \qquad \quad |N, \zeta \rangle &=& {\cal N}_{\mu,\nu, N} \left(\mu \hat{a}  + \nu \hat{a}^{\dagger} \right)^{N} |\zeta\rangle \nonumber \\
\hspace{-1.1cm} \qquad \quad            &=& {\cal N}_{\mu,\nu, N}\;\nu^N N! \sum^{N}_{k = 0} \left(\frac{\mu}{\nu}\right)^k\;\;\sum^{min(N-k, k)}_{l = 0}
\frac{\left(\frac{1}{2}\right)^l:\hat{a}^{\dagger^{N-k-l}}\hat{a}^{k-l}:}{l!\; (k - l)!\;(N-k-l)!}|\zeta\rangle,
\label{op4}
\end{eqnarray}
where ${\cal N}_{\mu,\nu, N}$ is the normalization constant. We fix the  constant, 
${\cal N}_{\mu,\nu, N}$, from the normalization condition $\langle N, \zeta| N, \zeta\rangle = 1$ 
which in turn  provides 
\begin{eqnarray}
\hspace{-0.6cm}\qquad {\cal N}_{\mu,\nu, N}^{-2} &=& |\nu|^{2 N} (N!)^2\sum^{N}_{k = 0} \left(\frac{|\mu|}{|\nu|}\right)^{2 k}
 \nonumber \\ 
\hspace{-0.6cm} \qquad & &\times\;\; \sum^{min(N-k, k)}_{l = 0} \frac{\left(\frac{1}{4}\right)^l\langle\zeta|:\hat{a}^{\dagger^{k-l}}\hat{a}^{N-k-l}: :\hat{a}^{\dagger^{N-k-l}}\hat{a}^{k-l}:|\zeta\rangle}{(l!\; (k - l)!\; (N-k-l)!)^2}.
\label{op5}
\end{eqnarray}

To evaluate the above expression (\ref{op5}) we insert the complete relation of the coherent states, 
$\frac{1}{\pi}\int^{\infty}_{-\infty}|\alpha\rangle \langle \alpha|\;d^2 \alpha = 1$,  so
that Eq. (\ref{op5})  becomes
\begin{eqnarray}
\hspace{-1cm}\qquad \qquad {\cal N}_{\mu,\nu, N}^{-2} &=& |\nu|^{2 N} (N!)^2 \sum^{N}_{k = 0} \left(\frac{|\mu|}{|\nu|}\right)^{2 k}\;\;
\sum^{min(N-k, k)}_{l = 0} \frac{\left(\frac{1}{4}\right)^l}{\pi(l!\;(k - l)!\;(N-k-l)!)^2}\nonumber \\
\hspace{-1cm}\qquad\qquad & & \times \int^{\infty}_{-\infty}\langle\zeta|:\hat{a}^{\dagger^{k-l}}\hat{a}^{N-k-l}:|\alpha\rangle \langle \alpha|:\hat{a}^{\dagger^{N-k-l}}\hat{a}^{k-l}:|\zeta\rangle\;d^2 \alpha.
\label{op6}
\end{eqnarray}
We first evaluate the integral (which we call it as $G$) appearing in (\ref{op6}). 
Evaluating the matrix elements of $\hat{a}^{\dagger^{k - l}}
\hat{a}^{N-k-l}$ and $\hat{a}^{\dagger^{N-k-l}} \hat{a}^{k-l}$ of the coherent states, we get
\begin{eqnarray}
\hspace{-1cm} \qquad \qquad \qquad G &=& e^{-\frac{|\zeta|^2}{2}} |\zeta|^{2 (k - l)} \int^{\infty}_{-\infty} e^{-|\alpha|^2 + \zeta^* \alpha + \zeta \alpha^*} |\alpha|^{2(N-k-l)} d^2 \alpha
\end{eqnarray}
which can be re-expressed  as
\begin{eqnarray}
\hspace{-1cm} \qquad \qquad \qquad G &=& e^{-\frac{|\zeta|^2}{2}} |\zeta|^{2 (k - l)} \frac{\partial^{2(N-k-l)}}{\partial \zeta^{N-k-l}\partial \zeta^{*^{N-k-l}}}\int^{\infty}_{-\infty} e^{-|\alpha|^2 + \zeta^* \alpha + \zeta \alpha^*} d^2 \alpha.
\label{op7}
\end{eqnarray}
Now employing the integral formula $\int^{\infty}_{-\infty} e^{-|z|^2 + a z + b z^*} d^2 z= \pi\; e^{a b}$, 
we find 
\begin{eqnarray}
\hspace{-1cm} \qquad \qquad \qquad G &=& \pi e^{-\frac{|\zeta|^2}{2}} |\zeta|^{2 (k - l)} \frac{\partial^{2(N-k-l)}}{\partial \zeta^{N-k-l}\partial \zeta^{*^{N-k-l}}}\left( e^{|\zeta|^2} \right)\nonumber\\
\hspace{-1cm} \qquad \qquad \qquad&=& \pi e^{-\frac{|\zeta|^2}{2}} |\zeta|^{2 (k - l)} \frac{\partial^{N-k-l}}{\partial \zeta^{N-k-l}}\left( e^{|\zeta|^2}\zeta^{N-k-l}\right) \nonumber \label{op8}\\
\hspace{-1cm} \qquad \qquad \qquad &=& \pi |\zeta|^{2 (k - l)} (N-k-l)! {\cal L}_{N-k-l} \left(-|\zeta|^2\right).
\label{op9a}
\end{eqnarray}
To obtain the last expression in (\ref{op9a}) we have used the Rodrigues' formula 
${\displaystyle {\cal L}_n (x) = \frac{e^x}{n!}\frac{\partial^n}{\partial x^n}(e^{-x} x^n)}$, where ${\cal L}_n(x)$ is the
Laguerre polynomial of order $n$.

Substituting (\ref{op9a}) in (\ref{op6}), we obtain the normalization constant for the
photon modulated coherent states $|N, \zeta\rangle$ of the form 
\small
\begin{eqnarray}
\hspace{-0.3cm}\qquad \qquad {\cal N}_{\mu,\nu, N}^{-2} &=& |\nu|^{2 N} (N!)^2 \sum^{N}_{k = 0} \left(\frac{|\mu|}{|\nu|}\right)^{2 k}\nonumber \\
\hspace{-0.3cm}\qquad \qquad                            & &\;\times\sum^{min(N-k, k)}_{l = 0} 
\frac{\left(\frac{1}{4}\right)^l|\zeta|^{2 (k - l)}\; (N-k-l)!\; {\cal L}_{N-k-l} \left(-|\zeta|^2\right)}{(l!\; (k - l)!\; (N-k-l)!)^2}.
\label{op10}
\end{eqnarray}
\normalsize 

In the case $N =0$ and $\mu = \nu = 0$, we find ${\cal N}_{\mu,\nu, N}^{-2} = 1$. When 
$\mu = 0$, $\nu =1$ and $N \neq 0$, we obtain 
\begin{equation}
\qquad \qquad {\cal N}_{0,1, N}^{-2} = N!\; {\cal L}_{N} \left(-|\zeta|^2\right), 
\label{op10a}
\end{equation}
which is the normalization constant of the photon added coherent state \cite{agarwal1}. 
When $\mu = 1$, $\nu = 0$ and $N \neq 0$, all the terms in the expression (\ref{op10}) 
vanish except $k = N$ which in turn yield $|\zeta|^{2 N}$. 
Our result confirms that the normalization constant (\ref{op10}) generalizes 
the normalization constant of the photon added coherent state (vide (\ref{op10a}))  and 
photon subtracted coherent state. 

\section{Photon statistical properties of the photon modulated coherent states}

In this section, we investigate the classical and non-classical properties of the
constructed photon modulated coherent states. For this purpose, we calculate the
quantities (i) $A_3$-parameter, (ii) quadrature squeezing and (iii)
$s$-parameterized quasi probability distribution function.

\subsection{$A_3$-parameter}
To test the non-classical character of the photon modulated coherent states $|N,  \zeta\rangle$,
we investigate the parameter $A_3$.  The parameter $A_3$ can be calculated from the
expression \cite{agar1},

\begin{eqnarray}
A_{3} = \frac{\det m^{(3)}}{\det\mu^{(3)} - \det m^{(3)}},
\label{A3}
\end{eqnarray}
where
\begin{eqnarray}
     m^{(3)}= \left(\begin{array}{ccc}
    1 & m_1 & m_2 \\
    m_1 & m_2 & m_3 \\
    m_2& m_3 & m_4\\
    \end{array} \right) \;\;\;\mbox{and}
\;\;\;\;\;&
  \mu^{(3)}= \left(
  \begin{array}{ccc}
   1 & \mu_1 & \mu_2 \\
    \mu_1 & \mu_2 & \mu_3 \\
    \mu_2& \mu_3 & \mu_4\\
    \end{array}
   \right). \label{mmu3}
\end{eqnarray}
In the expression (\ref{mmu3}), $m_j = \langle \hat{a}^{\dagger^j} \hat{a}^j \rangle $ and
$\mu_j = \langle (\hat{a}^{\dagger} \hat{a})^j \rangle$, $j = 1, 2, 3, 4 $.
This parameter was introduced as a counterpart to the Mandel's parameter $Q$.
To test the non-classical character of the field even if it does not exhibit the squeezing and
sub-Poissonian statistics, the Mandel's parameter $Q$ is generalized to a quantity $m^{(n)}$ formed
from the moments of Glauber-Sudarshan  function ($P$), $\hat{m}_n = \hat{b}^{\dagger^n} {\hat{b}}^n$,
where $\hat{b}$ and $\hat{b}^{\dagger}$ are the annihilation and creation operators of the harmonic oscillator.
To normalize $m^{(n)}$ another quantity $\mu^{(n)}$, where $\hat{\mu}^{(n)} = (\hat{b}^{\dagger} \hat{b})^n$,
which is formed from the moments of number distribution has also been introduced.
The normalized quantity with $n = 3$ obtained from $m^{(n)}$ and $\mu^{(n)}$ is termed as parameter $A_3$
(vide Eq. (\ref{A3})).

For the coherent and vacuum states $\det m^{(3)}=0$ and for a Fock state $\det m^{(3)}= -1$ and $\det\mu^{(3)} = 0$.
For the non-classical states $\det m^{(3)} < 0$ and since $\det\mu^{(3)} > 0$, it follows that the parameter $A_3$
lies between 0 and -1.

To calculate the parameter $A_3$, we  evaluate ${m}_j$'s and
${\mu}_j$'s, $j=1,2,3,4$, for the photon modulated coherent states $|N, \zeta\rangle$ as
\begin{eqnarray}
\hspace{-0.4cm} \;\; m_j  &=& {\cal N}_{\mu,\nu, N}^2 \frac{|\nu|^{2\;N}\; (N!)^2}{\pi^2}\;\sum^{N}_{k = 0} \left(\frac{|\mu|}{|\nu|}\right)^{2 k}\;\;
\sum^{min(N-k, k)}_{l = 0} \frac{\left(\frac{1}{4}\right)^l}{(l!\;(k - l)!\;(N-k-l)!)^2}\nonumber \\
\hspace{-0.4cm} \;\; & &\times\int^{\infty}_{-\infty}\int^{\infty}_{-\infty}\langle\zeta|:\hat{a}^{\dagger^{k-l}}\hat{a}^{N-k-l}:|\alpha\rangle \langle \alpha|:\hat{a}^{\dagger^j} \hat{a}^j:\beta\rangle\langle \beta|:\hat{a}^{\dagger^{N-k-l}}\hat{a}^{k-l}:|\zeta\rangle
 \nonumber \\
\hspace{-0.4cm} \;\; & & \hspace{8cm}\times \;d^{2}{\alpha}\;d^2 \beta.
\label{m3}
\end{eqnarray}

The double integral appearing in (\ref{m3}) can be evaluated in the same manner as in the 
case of finding the normalization constant. We skip this rather lengthy and straightforward 
derivation and the final form of the expression reads  
\begin{eqnarray}
\hspace{-1cm}\qquad \quad m_j  &=&  {\cal N}_{\mu,\nu, N}^2|\nu|^{2 N} (N!)^2 \sum^{N}_{k = 0} \left(\frac{|\mu|}{|\nu|}\right)^{2 k}
\sum^{min(N-k, k)}_{l = 0} \frac{\left(\frac{1}{4}\right)^l|\zeta|^{2 (k-l)}\; (N-k-l+j)!}{(l!\;(k - l)!\;(N-k-l)!)^2}\nonumber \\
\hspace{-1cm} \quad && \qquad\hspace{4.2cm}\times{\cal L}_{N-k-l+j}(-|\zeta|^2), \quad j = 1, 2, 3, 4.
\label{m4}
\end{eqnarray}

Equation (\ref{m4}) provides explicit expressions for the moments ($m_j$) 
of the Glauber-Sudarshan  function ($P$). 
 
To evaluate the expression $\mu_j = \langle (\hat{a}^{\dagger} \hat{a})^j \rangle$ we again express
the quantity $(\hat{a}^{\dagger} \hat{a})^j$ in  Weyl ordering \cite{agar}
\begin{equation}
(\hat{a}^{\dagger} \hat{a})^j_W = \sum^{j}_{i = 0}\sum^{i}_{r =  0} \frac{(-1)^r\;(i - r)^j}{r!\;(i - r)!}
:\hat{a}^{\dagger^j}\hat{a}^j :.
\label{mu1}
\end{equation}
Using  (\ref{mu1}), we find
\begin{eqnarray}
 \hspace{-0.2cm}\mu_j  &=&{\cal N}_{\mu,\nu, N}^2  |\nu|^{2 N} (N!)^2 \sum^{N}_{k = 0} \left(\frac{|\mu|}{|\nu|}\right)^{2 k}  \sum^{j}_{i = 0}\sum^{i}_{r =  0}\frac{(-1)^r\; (i - r)^j}{r!\; (i - r)!}\;\nonumber \\
 \hspace{-0.2cm} & & \; \times \sum^{min(N-k, k)}_{l = 0} \frac{\left(\frac{1}{4}\right)^l}{(l!\;(k - l)!\;(N-k-l)!\pi)^2}
\int^{\infty}_{-\infty}\;\int^{\infty}_{-\infty}\langle\zeta|:\hat{a}^{\dagger^{k-l}}\hat{a}^{N-k-l}:|\alpha\rangle \nonumber \\
\hspace{-0.2cm} & &\; \hspace{3cm} \times  \;\langle \alpha|:\hat{a}^{\dagger^i} \hat{a}^i:\beta\rangle\;\langle \beta|:\hat{a}^{\dagger^{N-k-l}}\hat{a}^{k-l}:|\zeta\rangle\;d^{2}{\alpha}\;d^2 \beta.
\label{mu2}
\end{eqnarray}

The double integral can   be evaluated in the same manner as we done previously. The final result shows
\begin{eqnarray}
 \hspace{-0.3cm}  \mu_j  &=&{\cal N}_{\mu,\nu, N}^2  |\nu|^{2 N} (N!)^2 \sum^{N}_{k = 0} \left(\frac{|\mu|}{|\nu|}\right)^{2 k}\;
 \sum^{j}_{i = 0}\sum^{i}_{r =  0} \frac{(-1)^r\;(i - r)^j}{r!\; (i - r)!}\; \nonumber \\
 \hspace{-0.3cm} & & \times \;\sum^{min(N-k, k)}_{l = 0}\frac{\left(\frac{1}{4}\right)^l
|\zeta|^{2 (k-l)}\;(N-k-l+i)!}{(l!\;(k - l)!\;(N-k-l)!)^2}\; {\cal L}_{N-k-l+i}(-|\zeta|^2), 
\label{mu3}
\end{eqnarray}
which can be used to calculate $\det \mu^{(3)}$. Using the expressions (\ref{m4}) and 
(\ref{mu3}) we can determine the parameter $A_3$. 

\begin{figure}[htP]
\centering
\includegraphics[width=0.9\linewidth]{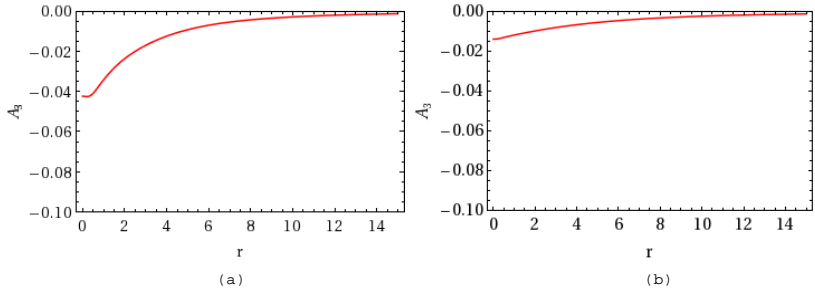}
\vspace{-0.1cm}
\caption{The plot of the parameter $A_3$ for
photon modulated coherent states $|N,\zeta\rangle$ with  $\mu = \frac{1}{3},\; \nu = \frac{2}{3}$ for different
values of $N$ as (a) $N = 2$ and (b) $N = 20.$ }
\label{a3}
\end{figure}

From the expectation values, we work out the parameter $A_3$,  for the states $|N,\zeta\rangle$
with $\zeta = r e^{i \theta}$, numerically.  We plot the results in figure \ref{a3}
where we have drawn the parameter $A_3$ against $r\;(= |\zeta|)$. The figures \ref{a3} (a) and (b) demonstrate
that the parameter $A_3$ lies in-between 0 to -1 which ensures  that the states given in (\ref{op4}) are
non-classical. Figure \ref{a3} also reveals the fact that 
for large values of $|\zeta|$ the parameter $A_3$
goes to zero. We also observe that the negativity of the parameter $A_3$ reduces 
while we increase the number $N$ (figure \ref{a3}(b)). 

\subsection{Quadrature squeezing}
The non-classical nature of a quantum state can also be confirmed by examining the degree of squeezing it
possesses. We analyze the squeezing in two new conjugate variables, namely deformed
position ($X$) and momentum ($Y$) coordinates, which are defined as \cite{walls}
\begin{eqnarray}
\hat{X} = \frac{1}{\sqrt{2}} (\hat{a}^{\dagger} + \hat{a}), \qquad \quad \hat{Y} = \frac{i}{\sqrt{2}} (\hat{a}^{\dagger} - \hat{a}).
\label{qua}
\end{eqnarray}

To analyze the squeezing in the quadratures $\hat{X}$ and $\hat{Y}$ in which
the Heisenberg uncertainty relation holds, $\left(\Delta \hat{X} \right)^2\left(\Delta \hat{Y} \right)^2 \ge \frac{1}{4}$,
where $\Delta \hat{X}$ and $\Delta \hat{Y}$ denote uncertainties in $\hat{X}$ and $\hat{Y}$
respectively, we evaluate the following two inequalities,
that is
\begin{eqnarray}
\quad I_{1} &=& \langle {\hat{a}}^2 \rangle + \langle \hat{a}^{\dagger^2} \rangle - \langle {\hat{a}} \rangle^{2} - \langle \hat{a}^{\dagger}\rangle^{2} - 2 \langle \hat{a} \rangle \langle \hat{a}^{\dagger} \rangle
+ 2\langle \hat{a}^{\dagger} \hat{a} \rangle < 0,
\label{id1}\\
\quad I_{2} &=& -\langle {\hat{a}}^2 \rangle - \langle \hat{a}^{\dagger^2} \rangle + \langle {\hat{a}} \rangle^{2} + \langle \hat{a}^{\dagger} \rangle^{2} - 2 \langle \hat{a} \rangle \langle \hat{a}^{\dagger} \rangle
+ 2\langle \hat{a}^{\dagger} \hat{a} \rangle < 0,
\label{id2}
\end{eqnarray}
which can be derived from the squeezing condition
$(\Delta \hat{X})^2 < \frac{1}{2}$ or $(\Delta \hat{Y})^2 < \frac{1}{2}$ by implementing
the expressions given in (\ref{qua}). The expectation values should be calculated with
respect to the photon modulated coherent states $|N, \zeta\rangle$ for
which the squeezing property has to be examined.

\begin{figure}[htP]
\centering
\includegraphics[width=0.45\linewidth]{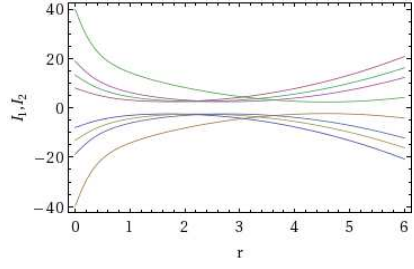}
\vspace{-0.1cm}
\caption{The plot of the identities for
photon modulated coherent states $|N,\zeta\rangle$ with  $\mu = \frac{1}{3},\; \nu = \frac{2}{3}$ and $N = 1,2,3,6$ }
\label{sqp}
\end{figure}

The identities $I_1$ and $I_2$ are calculated numerically and plotted in figure \ref{sqp}.
We have drawn the  curves for $N = 1,2,3,6$ with $\mu = \frac{1}{3}$ and $\nu = \frac{2}{3}$.
The curves  above and below $0$ are associated with $I_1$ and $I_2$ respectively.
The squeezing in $Y$ can be observed for all values of $\mu, \nu $ and $N$ which is
clearly illustrated in figure \ref{sqp}. From our analysis we
conclude  that the states $|N, \zeta\rangle$ are non-classical.

\subsection{$s$-parameterized quasi-probability function}
We also confirm the non-classicality nature of the photon modulated coherent states
by studying the $s$-parameterized quasi-probability distribution function for the
states (\ref{op4}), with $s$ is a continuous variable.
The $s$-parameterized function was introduced by Cachill and Glauber as the generalized function
that interpolates the Glauber-Sudarhsan $P$-function for $s$ = 1, Wigner function $W$ for $s = 0$
and Husimi $Q$-function for $s = -1$ \cite{cah}.

The $s$-parameterized quasi-probability
distribution function, $F(\gamma, s)$, is the
Fourier transform of the $s$-parameterized characteristic function \cite{cah, vbook, barn}
\begin{eqnarray}
\hspace{-1cm} \qquad \qquad \qquad F(\gamma, s) = Tr[\hat{\rho} \hat{T}(\gamma, s)],
\label{spara}
\end{eqnarray}
where
\begin{eqnarray}
\hspace{-1cm} \qquad \qquad \qquad \hat{T}(\gamma, s) = \frac{2}{1-s}\hat{D}(\gamma)\exp{\left(\ln\frac{1+s}{s-1} \hat{a}^{\dagger}\hat{a}\right)}
\hat{D}^{-1}(\gamma), 
\label{char}
\end{eqnarray}
is the $s$-parameterized characteristic function operator \cite{obada} and $\hat{D}(\gamma)$ is the
displacement operator.

We express the operator function $\exp{\left(\lambda \hat{a}^{\dagger}\hat{a}\right)}$,
where $\lambda = \ln\left(\frac{s+1}{s-1}\right)$,  appearing in
Eq. (\ref{char}), in terms of Weyl ordered form through the identity
\begin{equation}
\hspace{-1cm} \qquad \qquad e^{-\lambda \hat{a}^{\dagger}\hat{a}} = \frac{2}{1+e^{-\lambda}} \exp{\left( \frac{-2(1-e^{-\lambda})}{1+e^{-\lambda}} :\hat{a}^{\dagger}\hat{a}:\right)}.
\label{eop}
\end{equation}

To obtain the $s$-parameterized quasi-probability distribution function for the photon modulated
coherent states, $\hat{\rho} = |N, \zeta\rangle\langle N, \zeta|$,
we first determine the  operator $\hat{T}(\gamma,\;s)$ by substituting the 
expression (\ref{eop}) in (\ref{char}). We then insert  the 
obtained expression in (\ref{spara}). Doing so we find
\small
\begin{eqnarray}
\hspace{-1cm} \qquad \quad F(\gamma, s) &=& \frac{2\;{\cal N}_{\mu,\nu, N}^2\;|\nu|^{2 N}\;(N!)^2}{\pi^4\;(1-s)}\;\;\sum^{N}_{k = 0}\;\left(\frac{|\mu|}{|\nu|}\right)^{2 k}
\sum^{min(N-k, k)}_{l = 0} \frac{\left(\frac{1}{4}\right)^l}{(l!\;(k - l)!\;(N-k-l)!)^2}\nonumber \\
\hspace{-1cm} \qquad \quad & &\times \oint\oint\oint\oint \langle\zeta|\hat{G}^{\dagger}_N|\alpha\rangle\;\langle \alpha|D(\gamma)|\beta\rangle\;
\langle |\beta|\exp{\left(-2 \frac{1-e^{-\lambda}}{1+e^{-\lambda}} :\hat{a}^{\dagger}\hat{a}:\right)}|\xi\rangle\nonumber \\
\hspace{-1cm} \qquad \quad & &\qquad \qquad \times  \langle\xi| D^{-1}(\gamma)|\eta\rangle\;\langle\eta|\hat{G}_N|\zeta\rangle\; d^2\alpha\; d^2\beta\;
d^2\xi d^2\eta,
\label{char1}
\end{eqnarray}
\normalsize
Equation (\ref{char1}) has been obtained by taking into account the completeness relation 
$\frac{1}{\pi}\oint |\alpha\rangle\langle\alpha|\;d^2\alpha = 1$ four times. 

As usual, we first evaluate the integrals appearing in (\ref{char1}). After a very lengthy calculation we
arrive at 
\begin{eqnarray}
\hspace{-0.5cm} \quad\hat{F}(\gamma, s) &=& \frac{2\;{\cal N}_{\mu,\nu, N}^2\;|\nu|^{2 N}\;(N!)^2}{\pi^2\;(1-s)}\;\;\exp{\left[- \frac{2+s}{s} (|\gamma|^2 + |\zeta|^2) + \left(\frac{s+1}{s}\right)(\gamma^*\zeta + \gamma \zeta^*)\right]}\nonumber \\
\hspace{-0.5cm} \quad & & \times \sum^{N}_{k = 0} \left(\frac{|\mu|}{|\nu|}\right)^{2 k}\;\;\;
\sum^{min(N-k, k)}_{l = 0} \frac{\left(\frac{1}{4}\right)^l|\zeta|^{2(k-l)}\;(N-k-l)!}{(l!\; (k - l)!\;(N-k-l)!)^2}\;\left(\frac{s-2}{s}\right)^{N-k-l} \nonumber \\
\hspace{-0.5cm} \quad & &\;\;\;\times{\cal L}_{N-k-l}\;\left[\frac{2 |\gamma|^2}{s+2} - \frac{2+s}{s} |\gamma|^2 + \frac{s+1}{s}\;(\gamma^*\zeta + \gamma \zeta^*)\right],
\label{spara_v}
\end{eqnarray}
where ${\cal L}_n$ is the Laguerre polynomial of order $n$.

\begin{figure}[htP]
\centering
\includegraphics[width=0.9\linewidth]{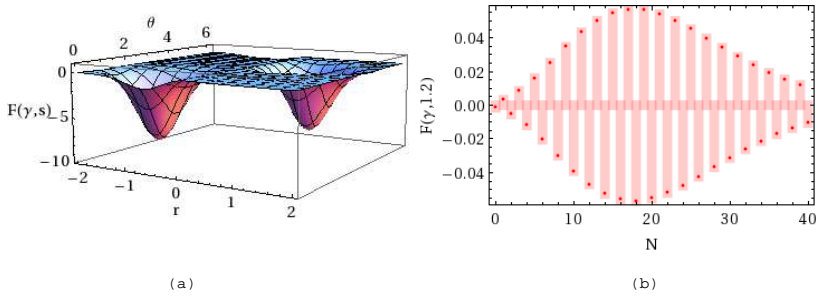}
\vspace{-0.1cm}
\caption{The plot of the quasi distribution function $F(\gamma, s)$ for
photon modulated coherent states $|N,\zeta\rangle$ with (a)  $s=1.2,\;\mu= 0.001,\nu= 1.2$ and $N = 2$ and
(b) $s =1.2,\; \mu = 0.001,\;\nu= 1.2,\;\zeta=  -1 e^{i\; 0.1}$ and  $\gamma = i$ .  }
\label{sp}
\end{figure}

Using the expression (\ref{spara_v}), we determine the $s$-parameterized quasi-probability
distribution function numerically for two sets of values, namely
(i) $s=1.2,\;\mu= 0.001,\;\nu= 1.2,\; \zeta = i$, $N = 2$ with
$\gamma = r e^{i \theta}$ and (ii) $s =1.2,\; \mu = 0.001,\;\nu= 1.2,\;\zeta=  -1 e^{i\; 0.1}$
with  $\gamma = i $ and the resultant function is evaluated for
different values of $N$. The results obtained  from both the cases are plotted in figures \ref{sp}(a) and (b) 
respectively. The non-classicality of the states $|N, \zeta\rangle$ can be seen explicitly in figure \ref{sp}(a) with
the negativity of $F(\gamma, s)$. In figure \ref{sp}(b) we depict the values of $s$-parameterized quasi-probability distribution function for
various values of $N$. The figure shows that the function $F(\gamma,s)$ fluctuates  between positive and negative
values for varying $N$ and saturate at large value of $N$. However, the negativity of
$F(\gamma, 1.2)$ unambiguously confirms the non-classicality of the photon modulated coherent states.

\subsection{Fidelity}
In this sub-section we examine the fidelity between modulated coherent states and 
the original coherent states. The fidelity is a non-Gaussianity measure able to 
quantify the non-Gaussian character of a quantum state \cite{fidelity}. Using this we confirm the 
non-classicality of the constructed state. 

To calculate the fidelity,  we evaluate  
\begin{eqnarray}
F_{\mu, \nu, N} &=& \frac{|\langle\zeta|N, \zeta\rangle|^2}{|\langle \zeta|\zeta\rangle|^2}.  
\end{eqnarray}
Substituting (\ref{op4}) in the above definition with $|\langle \zeta|\zeta \rangle|^2 = 1$ 
and evaluating  the resultant expression, we arrive at the following expression for the fidelity 
between photon modulated coherent state and its coherent state, namely 
\begin{eqnarray}
 \hspace{-0.5cm} \quad \quad  F_{\mu, \nu, N}  &=&  {\cal N}_{\mu,\nu, N}^2\;|\nu|^{2\;N}\;(N!)^2\;\nonumber \\
                                               & & \times\sum^{N}_{k = 0} \left(\frac{|\mu|}{|\nu|}\right)^{2 k}\;
                                                    \sum^{min(N-k, k)}_{l = 0}\frac{\left(\frac{1}{4}\right)^l
                                                   |\zeta|^{2\;(N- 2l)}}{(l!\;(k - l)!\;(N-k-l)!)^2}. 
\label{fidf}
\end{eqnarray}

In the limit $N = 0$, we should get $ F_{\mu, \nu, 0} = 1$ which can be directly verified from (\ref{fidf}). 

\begin{figure}[htP]
\centering
\includegraphics[width=0.4\linewidth]{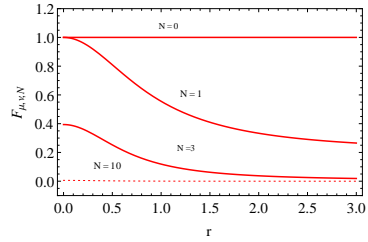}
\vspace{-0.1cm}
\caption{The plot of fidelity of
photon modulated coherent states $|N,\zeta\rangle$ with  $\mu = \frac{1}{3},\; \nu = \frac{2}{3}$ for different
values of $N$ as (a) $N = 0, 1, 3$ and $10$. }
\label{fid}
\end{figure}

We plot the fidelity (\ref{fidf}) between photon modulated coherent state and its 
original state as a function of $|\zeta| = r$ for four different values of $N$
say $N = 0, 1, 2$ and $10$. We fix the parameters $\mu = \frac{1}{3}$ and $\nu = \frac{2}{3}$. 
For $N = 0$, photon modulated coherent state reduces to
the coherent state which is again confirmed by getting a straight line at $F_{\mu, \nu, N}  = 1$.
The fidelity should decrease when we increase the number of photons. This has
also been clearly illustrated in figure \ref{fid}. This test also validates the non-classicality of the states
$|N, \zeta\rangle$.

\section{Conclusion}
In this paper, we have constructed photon modulated coherent states 
of a  generalized isotonic oscillator potential (\ref{pot}) by considering a generalized
superposed operator $(\mu \hat{a} + \nu \hat{a}^{\dagger})^N$. We implemented   
Weyl ordering to expand this generalized operator. The resultant expression comes out with two parameter Hermite polynomial,
${\cal H}_{m,n}(\hat{a}, \hat{a}^{\dagger})$, with a 
finite series in $:\hat{a}^{\dagger^m} \hat{a}^n:$. 
Since we have already known the ladder operators, we plugged them in the superposed
operator $(\mu \hat{a} + \nu \hat{a}^{\dagger})^N$ and obtained the 
photon modulated coherent states of the potential under investigation. 
We have evaluated the parameter $A_3$,
quadrature squeezing identities $I_1$ and $I_2$ and the $s$-parameterized function for the
constructed photon modulated coherent states. All our results confirm
the non-classical nature of the photon modulated  coherent states. Since the coherent states of the
generalized isotonic oscillator is Gaussian with respect to new canonical variable $X$ \cite{chi3}, we have
also analyzed the transition from Gaussianity  to non-Gaussianity by evaluating the
fidelity between photon modulated coherent states and coherent states. This result also  confirms the
non-classicality nature of the photon modulated coherent states. The conclusions present in this 
paper will add further knowledge on this potential.  

\section*{Acknowledgements}
VC  wishes to thank the Council of Scientific and Industrial Research,
Government of India for providing a Senior Research Fellowship.

\section*{}

\end{document}